\documentclass[lineno]{jfm}

\usepackage{graphicx}
\usepackage{indentfirst}
\usepackage{float}
\usepackage{subfigure}
\usepackage{newtxtext}
\usepackage{newtxmath}
\usepackage{natbib}
\usepackage{hyperref}
\usepackage{orcidlink}
\hypersetup{
    colorlinks = true,
    urlcolor   = red,
    citecolor  = blue,
}

\newcommand{\RomanNumeralCaps}[1]

\title{\nolinenumbers Theoretical analysis for non-linear effects of magnetic fields on unsteady boundary layer flows}

\author{\nolinenumbers Jing-Yu Fu\aff{1},
  Ming-Jiu Ni\aff{1}
 \and Nian-Mei Zhang\aff{1}\corresp{\email{nmzhang@ucas.ac.cn}}}

\affiliation{\nolinenumbers \aff{1}School of Engineering Science, University of Chinese Academy of Sciences, Beijing 101408, PR China}

\begin{document}

\nolinenumbers\maketitle\nolinenumbers

\begin{abstract}
This study investigates unsteady boundary layer phenomena in electrically conducting fluids subjected to static magnetic fields. Using a semi-explicit similarity transformation method, the momentum equation associated with the Stokes stream function is solved. The nonlinear closed analytical solutions for both stagnation flow and converging flow are derived. The results demonstrate that the boundary layer structure incorporates shock and solitary wave-like components, which are promoted by the Lorentz force. Under extreme magnetic fields, the flow exhibits sine and cosine wave patterns, which are motivated by the strong Lorentz force. An in-depth asymptotic analysis establishes the square root scaling laws that quantify the growth of friction and flux with increasing magnetic field strength. The boundary layer thickness scales inversely with the Hartmann number, a consequence of the dominant Lorentz force, which differs from the conclusion for duct flow \citep{Hunt1965}. These findings elucidate the physical mechanisms governing the nonlinear coupling between magnetic fields and the dynamics of the boundary layer.
\end{abstract}


\section{Introduction}
\label{sec:headings}

Boundary layer flow is a classical problem in fluid mechanics, introduced by \cite{Prandtl1904}. Mathematically, its governing equation is the Navier-Stokes equation, which has been approximated at high Reynolds numbers \citep{Landau1976}. However, it has exact solutions only under certain special assumptions \citep{Wang1991}. For example, the simplest assumption is a steady and zero pressure gradient \citep{Blasius2024}. The study of the unsteady boundary layer has broader physical and engineering significance. Most studies of the unsteady boundary layer used numerical simulations and experimental measurements, with few analytical solutions derived mathematically. The power series solution of unsteady boundary layer flow in a given free stream velocity condition was obtained by \cite{Hassan1960}. Similarity transformation is an efficient mathematical approach for finding solutions to partial differential equations in fluid mechanics \citep{Ungarish2024}. \cite{Sun2024} proposed a diffusion time scale similarity transformation method and derived analytical solutions for two types of unsteady boundary layers.

The balance between viscous and inertial forces in the boundary layer is disturbed when other body forces are present in the fluid. In magnetohydrodynamics (MHD) \citep{Fu2025}, this body force, also known as the Lorentz force, inevitably affects the dynamic properties of the boundary layer. \cite{Zhang2011} reduced the steady MHD convergence boundary layer to the Falkner-Skan equation and derived an analytical solution. For the unsteady MHD boundary layer problem, \cite{Takhar1997} provides a numerical solution of the ordinary differential equation for axisymmetric stagnation flow based on Ma's Lie group method \citep{Ma1990}. Few scholars have explored analytical solutions for unsteady MHD boundary layer flow. Therefore, there is a lack of a precise theoretical explanation for the strong magnetic field's influence on the boundary layer's physical parameters.

The second section proposes a Stokes stream function similarity transformation method applicable to multiple flow types. The third section derives analytical solutions for the unsteady MHD boundary layer and, based on these analytical solutions, investigates the scaling relationship between the significant magnetic field strength and the velocity or its gradient. The fourth section extends the analytical solution to the boundary layer issue of converging flow under a non-uniform magnetic field. The collective analysis of these sections will elucidate the physical mechanism underlying the influence of strong magnetic fields on boundary layer flow. 

\section{Dynamic Equation}\label{notstyle}

\begin{figure}
  \centerline{\includegraphics[width=0.7\textwidth]{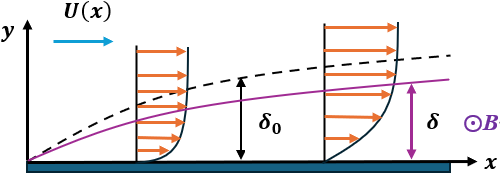}}
  \caption{Schematic diagram of two-dimensional laminar boundary layer under static magnetic field.}
\label{fig1}
\end{figure}

A model of two-dimensional Magnetohydrodynamic unsteady boundary layer flow is formulated. A Cartesian coordinate system $o-xyz$ is established, wherein the flow is confined to the $x-y$ plane as illustrated in figure \ref{fig1}. At time $t$, $u$ and $v$ represent the tangential (parallel to wall) and normal (perpendicular to wall) velocities of the fluid, $p$ pressure, and $\phi$ electrical potential. The Navier-Stokes equation in laminar momentum boundary layer for the viscous incompressible conductive fluid with Lorentz force and mass continuity equation is written as follows under a spanwise magnetic field $\boldsymbol{B}=B\boldsymbol{e_z}$
\begin{equation}
  \rho(\frac{\partial u}{\partial t}+u\frac{\partial u}{\partial x}+v\frac{\partial u}{\partial y})=-\frac{\partial p}{\partial x}+\mu \frac{\partial^2u}{\partial y^2}-\sigma B^2u-\sigma B\frac{\partial \phi}{\partial y}
\end{equation}
\begin{equation}
  \frac{\partial u}{\partial x}+\frac{\partial v}{\partial y}=0.
\end{equation}
\begin{equation}
  \frac{\partial p}{\partial y}=0
\end{equation}
where $\sigma$, $\mu$ and $\rho$ denote the conductivity, viscosity and density of the fluid, respectively.

The validity of the aforementioned three equations is contingent upon the large Reynolds number condition, a prerequisite for the neglect of \textit{O}($1/\Rey$) terms, as postulated by Prandtl theory \citep{Landau1976}. The region outside the boundary layer can be regarded as an inviscid potential flow (main flow), therefore the dynamic governing equation for the main flow velocity $U$ is simplified as
\begin{equation}
  \rho U\frac{\partial U}{\partial x}=-\frac{\partial P}{\partial x}-\sigma B^2U-\sigma B\frac{\partial \phi}{\partial y}
\end{equation}

It can be inferred from the zero normal pressure gradient that $p=P$. Assuming that the solid wall surface is electrically insulated, this boundary condition will result in the electrical potential gradient term in momentum equations accounting for half of the total Lorentz force \citep{Fu2025}. Consequently, the tangential N-S equation within the boundary layer is expressed as follows
\begin{equation}
  \rho(\frac{\partial u}{\partial t}+u\frac{\partial u}{\partial x}+v\frac{\partial u}{\partial y})=\rho U\frac{\partial U}{\partial x}+2\sigma B^2U+\mu \frac{\partial^2u}{\partial y^2}-2\sigma B^2u
\end{equation}

The two-dimensional continuity equation enables the introduction of the Stokes stream function, denoted by $\psi$, to rewrite the momentum equation
\begin{equation}
  \rho(\frac{\partial^2 \psi}{\partial t\partial y}+\frac{\partial \psi}{\partial y}\frac{\partial^2 \psi}{\partial x\partial y}-\frac{\partial \psi}{\partial x}\frac{\partial^2 \psi}{\partial^2 y})=\rho U\frac{\partial U}{\partial x}+2\sigma B^2U+\mu \frac{\partial^3\psi}{\partial y^3}-2\sigma B^2\frac{\partial \psi}{\partial y}
\end{equation}
Here, the time scale of viscous dissipation within the boundary layer is employed to perform a similarity transformation on the above momentum equation \citep{Sun2024}. The dimensionless time, normal coordinate, and stream function corrected by the MHD effect are, respectively, as follows
\begin{equation}
  \tau=\frac{\mu t}{\rho \delta^2}
\end{equation}
\begin{equation}
  \eta=\frac{y}{\delta}
\end{equation}
\begin{equation}
  \psi=f(\tau,\eta)\delta(x)U(x)=[g(\tau,\eta)-f_L(\tau,\eta)]\delta(x)U(x)
\end{equation}

In above formulas, the $f_L$ is a linear function with an explicit magnetic field strength parameter, while $g$ is a nonlinear function. The $\delta$ represents the magnitude of boundary layer thickness while the $\delta_0$ in figure \ref{fig1} represents the case without magnetic field. The momentum equation of the boundary layer after this similarity transformation is
\begin{equation}
  \frac{\partial^3f}{\partial\eta^3}-\frac{\partial^2f}{\partial\eta\partial\tau}-\beta(\frac{\partial f}{\partial\eta})^2+\alpha f\frac{\partial^2f}{\partial\eta^2}+1=2\frac{\sigma B^2}{\mu}\delta^2(\frac{\partial f}{\partial\eta}-1)+\gamma\tau(\frac{\partial f}{\partial\tau}\frac{\partial^2f}{\partial\eta^2}-\frac{\partial f}{\partial\eta}\frac{\partial^2f}{\partial\eta\partial\tau})
\end{equation}

The self similarity condition requires that the coefficients of each term do not explicitly include the tangential coordinate $x$. Consequently, $\delta^2=\frac{\mu}{\rho}\frac{\lvert\beta-2\alpha\rvert x}{U(x)}$, $U(x)=Wx^{k}(k\in Z)$, $\beta=Re_{\delta}\frac{dU/dx}{u_{\delta}/\delta}$, $\alpha=Re_{\delta}(\frac{dU/dx}{u_{\delta}/\delta}+\frac{d\delta/dx}{u_{\delta}/U})$, $\gamma=2Re_{\delta}\frac{d\delta/dx}{u_{\delta}/U}$. Where $Re_{\delta}=\frac{\rho\delta u_{\delta}}{\mu}$ is the Reynolds number in the boundary layer. These similarity coefficients $\beta$, $\alpha$, $\gamma$ characterize the influence of external pressure on flow strength in the boundary layer. A comprehensive investigation of the influence of the magnetic field on the stagnation flow and converging flow boundary layers is elucidated in the following sections.

\section{Stagnation Flow $(k=1)$}\label{Stagnation}
In a uniformly distributed magnetic field, the principle of similarity can be fully upheld only when the boundary layer thickness $\delta$ remains constant. Otherwise, the coefficient of the Lorentz force term will contain the $x$-coordinate. This section aims to derive an analytical solution for this pattern. It is not difficult to determine that the velocity in the main flow region $U$ must be a linear function of the tangential coordinate $x$, implying $k=1$. This phenomenon is referred to as stagnation point flow \citep{Fang2019}. Consequently, the momentum equation is simplified as
\begin{equation}
  \frac{\partial^3f}{\partial\eta^3}-\frac{\partial^2f}{\partial\eta\partial\tau}-\beta(\frac{\partial f}{\partial\eta})^2+\beta f\frac{\partial^2f}{\partial\eta^2}+\beta=2\beta\frac{\sigma B^2}{\rho W}(\frac{\partial f}{\partial\eta}-1)
\end{equation}

This text will first present an analytical solution for a nonlinear partial differential equation applicable to various multi-physical field parameters. However, in extreme cases, where there is a significant difference between the pressure and the Lorentz force, this non-linear equation can be simplified to a linear equation. The analytical solution in this scenario will provide deeper insight into the spatiotemporal characteristics of the boundary layer.

\subsection{General Solution under Arbitrary Magnetic Field}
Performing a two-order Taylor expansion of the two nonlinear terms in the momentum equation near the boundary, in conjunction with the non-penetration and non-slip boundary conditions, results in the following equations
\begin{equation}
  f\frac{\partial^2f}{\partial\eta^2}=(f\vert_{\eta=0}+\frac{\partial f}{\partial\eta}\vert_{\eta=0}d+\frac{1}{2}\frac{\partial^2f}{\partial\eta^2}d^2)\frac{\partial^2f}{\partial\eta^2}=\frac{1}{2}(\frac{\partial^2f}{\partial\eta^2})^2d^2
\end{equation}
\begin{equation}
  (\frac{\partial f}{\partial\eta})^2=(\frac{\partial f}{\partial\eta}\vert_{\eta=0}+\frac{\partial^2f}{\partial\eta^2}d)^2=(\frac{\partial^2f}{\partial\eta^2})^2d^2
\end{equation}

Where $d$ is a geometric small quantity relative to the boundary layer scale. So two nonlinear terms can be reduced to a single square term within the region near the wall. For the sake of convenience in the derivation process, it is reasonable to set $\beta=1$. It is noteworthy that $\frac{\sigma B^2}{\rho W}$ corresponds to the Stuart number $N$ \citep{Turkyilmazoglu2012}. The momentum equation can be further simplified as follows
\begin{equation}
  \frac{\partial^3f}{\partial\eta^3}-\frac{\partial^2f}{\partial\eta\partial\tau}-\frac{1}{2}(\frac{\partial f}{\partial\eta})^2+1=2N(\frac{\partial f}{\partial\eta}-1)
\end{equation}

Based on the idea of substitution, taking $2N$ as the proportional coefficient of $f_L$, the final simplified form of momentum equation can be obtained
\begin{equation}
  \frac{\partial^3g}{\partial\eta^3}-\frac{\partial^2g}{\partial\eta\partial\tau}-\frac{1}{2}(\frac{\partial g}{\partial\eta})^2+1+2N+2N^2=0
\end{equation}

Where nonlinear function $g=f+2N\eta$. The mathematical form of this partial differential equation is analogous to the Falkner-Skan equation \citep{Zhang2011}, with the distinction being the time-varying term. A closed analytical solution can be expressed as
\begin{equation}
  g(\eta, \tau)=c_{3b}\tanh(c_{1b}\eta+c_{2b}\tau+c_b)-c_{4b}\ln\cosh(c_{1b}\eta+c_{2b}\tau+c_b)+h(\tau)
\end{equation}
\begin{equation}
  h(\tau)=-c_{3b}\tanh(c_{2b}\tau+c_b)+c_{4b}\ln\cosh(c_{2b}\tau+c_b)
\end{equation}

By thoroughly matching the powers of tanh, exact solutions for each coefficient of every term can be easily obtained. It is important to note that the only parameter currently undetermined is $c_b$.
\begin{equation}
  f(\eta, \tau)=-2N\eta-\sqrt{6\sqrt{2}\sqrt{1+2N+2N^2}}\tanh\xi+2\sqrt{6\sqrt{2}\sqrt{1+2N+2N^2}}\ln\cosh\xi+h(\tau)
\end{equation}
\begin{equation}
  \xi=\frac{1}{2}\sqrt{\frac{\sqrt{2}\sqrt{1+2N+2N^2}}{6}}\eta+\frac{5}{12}\sqrt{2}\sqrt{1+2N+2N^2}\tau+c_b
\end{equation}
\begin{equation}
  h(\tau)=\sqrt{6\sqrt{2}\sqrt{1+2N+2N^2}}\tanh\tau_1-2\sqrt{6\sqrt{2}\sqrt{1+2N+2N^2}}\ln\cosh\tau_1
\end{equation}
\begin{equation}
  \tau_1=\frac{5}{12}\sqrt{2}\sqrt{1+2N+2N^2}\tau+c_b
\end{equation}

It is evident that this solution inherently fulfills the boundary condition for normal velocity. However, to determine the value of $c_b$, it is necessary to apply the boundary condition for tangential velocity (non-slip). It is important to note that the current analytical solution is only valid for a short-time approximation \citep{Sun2024}. The accuracy of the solution significantly improves with an increase in the strength of the magnetic field.
\begin{equation}
  \frac{\partial f}{\partial\eta}\vert_{\eta=0}=c_{3b}c_{1b}[1-(\tanh c_b)^2]-c_{4b}c_{1b}\tanh c_b-2N=0
\end{equation}
\begin{equation}
  c_b=\tanh^{-1}\frac{c_{4b}c_{1b}+\sqrt{(c_{4b}c_{1b})^2+4c_{3b}c_{1b}(c_{3b}c_{1b}-2S)}}{-2c_{3b}c_{1b}}
\end{equation}

This derivative of $f$ demonstrates that the boundary layer flow velocity, when subjected to an external magnetic field, retains a structure in which shock wave-like ($\tanh{\xi}$) and solitary wave-like ($1/\cosh^{2}{\xi}$) coexist. The Lorentz force plays a significant role in increasing their amplitude, frequency and wavenumber.

\subsection{Linear Solution under Strong Magnetic Field}
Under extreme physical conditions, the Lorentz force will excite new function forms in the boundary layer flow structure. If the similarity coefficient is quite small and the Stuart number is sufficiently large, the nonlinear terms in the momentum equation can be ignored. Under these extreme conditions, the momentum equation simplifies to a linear equation
\begin{equation}
  \frac{\partial^3f}{\partial\eta^3}-\frac{\partial^2f}{\partial\eta\partial\tau}=2\beta\frac{\sigma B^2}{\rho W}(\frac{\partial f}{\partial\eta}-1)
\end{equation}

The order of the spatial derivative in this equation can be simplified, and the integral residue is an unknown function of time.
\begin{equation}
  \frac{\partial^2f}{\partial\eta^2}-\frac{\partial f}{\partial\tau}-2\beta\frac{\sigma B^2}{\rho W}f+2\beta\frac{\sigma B^2}{\rho W}\eta=D(\tau)
\end{equation}

Let an auxiliary function $E(\tau)$ about time satisfy $\frac{\mathrm{d}E}{\mathrm{d}\tau}+2\beta\frac{\sigma B^2}{\rho W}E=D$. Then the equation is transformed into homogeneous form
\begin{equation}
  \frac{\partial^2}{\partial\eta^2}(f-\eta+E)-\frac{\partial}{\partial\tau}(f-\eta+E)-2\beta\frac{\sigma B^2}{\rho W}(f-\eta+E)=0
\end{equation}

Let function $(f-\eta+C)$ be extended to $F$ in the complex field. The method of separation of space and time variables can be applied to the derivation of an analytical solution to this equation. The variable $g_n$ represents the spatial function and $T_n$ represents the temporal function. The infinite series expansion is
\begin{equation}
  F=\sum_{n=0}^{\infty}g_n(\eta)T_n(\tau)
\end{equation}

A thorough examination of the characteristic root equations reveals the following conclusions
\begin{equation}
  F=\sum_{n=0}^{\infty}F_{0n}\exp{(-\eta\sqrt{2\beta\frac{\sigma B^2}{\rho W}+\mathrm{i}\frac{(2n+1)\pi}{2}})}[\cos{\frac{(2n+1)\pi}{2}\tau}+\mathrm{i}\sin{\frac{(2n+1)\pi}{2}\tau}]
\end{equation}

Utilizing the non-slip BC at the wall
\begin{equation}
  \frac{\partial f}{\partial\eta}\vert_{\eta=0}=1-\Real(\sum_{n=0}^{\infty}F_{0n}\sqrt{2\beta\frac{\sigma B^2}{\rho W}+\mathrm{i}\frac{(2n+1)\pi}{2}}[\cos{\frac{(2n+1)\pi}{2}\tau}+\mathrm{i}\sin{\frac{(2n+1)\pi}{2}\tau}])=0
\end{equation}
The Fourier expansion of 1 is readily obtained as $1=\frac{4}{\pi}\sum_{n=0}^{\infty}\frac{(-1)^n}{2n+1}\cos{\frac{(2n+1)\pi}{2}\tau}$. Subsequent to this, the undetermined coefficient $F_{0n}$ is obtained
\begin{equation}
  F_{0n}=\frac{4(-1)^n}{(2n+1)\pi\sqrt{2\beta\frac{\sigma B^2}{\rho W}+\mathrm{i}\frac{(2n+1)\pi}{2}}}
\end{equation}

The initial step is to ascertain the square root of the complex number. For the sake of clarity and concision in subsequent equations, it is expedient to introduce a parameter $R$ to represent the real part of this square root.
\begin{equation}
  R=\sqrt{\beta\frac{\sigma B^2}{\rho W}+\sqrt{(\beta\frac{\sigma B^2}{\rho W})^2+\frac{(2n+1)^2\pi^2}{16}}}
\end{equation}

Utilizing the non-penetrable BC at the wall
\begin{equation}
  E(\tau)=\sum_{n=0}^{\infty}\frac{4(-1)^n}{(2n+1)\pi[R^2+\frac{(2n+1)^2\pi^2}{16R^2}]}[R\cos{\frac{(2n+1)\pi}{2}\tau}+\frac{(2n+1)\pi}{4R}\sin{\frac{(2n+1)\pi}{2}\tau}]
\end{equation}
The analytical solution to the linear equation is subsequently obtained. To differentiate it from the nonlinear solution, it is written as $f_R$
\begin{equation}
  f_R=\eta -E(\tau)+\sum_{n=0}^{\infty}\frac{4(-1)^n}{(2n+1)\pi[R^2+\frac{(2n+1)^2\pi^2}{16R^2}]}\exp{(-R\eta)}(f_1+f_2+f_3+f_4)
\end{equation}
\begin{equation}
  f_1=R\cos{\frac{(2n+1)\pi}{4R}\eta}\cos{\frac{(2n+1)\pi}{2}\tau}
\end{equation}
\begin{equation}
  f_2=R\sin{\frac{(2n+1)\pi}{4R}\eta}\sin{\frac{(2n+1)\pi}{2}\tau}
\end{equation}
\begin{equation}
  f_3=-\frac{(2n+1)\pi}{4R}\sin{\frac{(2n+1)\pi}{4R}\eta}\cos{\frac{(2n+1)\pi}{2}\tau}
\end{equation}
\begin{equation}
  f_4=\frac{(2n+1)\pi}{4R}\cos{\frac{(2n+1)\pi}{4R}\eta}\sin{\frac{(2n+1)\pi}{2}\tau}
\end{equation}

Under sufficiently strong magnetic fields, the pressure parameter $\beta$ becomes negligible in magnitude and leads to the dominance of the linear terms in equation (3.1). Consequently, solitary and shock wave-like disappear, giving way to harmonic wave solutions expressible in terms of trigonometric functions. The strong magnetic field not only stimulates trigonometric waves, but it also decreases their amplitudes. This further illustrates the significant interaction between the magnetic field and the boundary layer flow.

\subsection{Coupling Effects of Magnetic Field and Flow}
The most concerning physical problem is how an external magnetic field impacts the flow characteristics within the boundary layer. In this section, the velocity profile, the damping force acting on the wall, flux of the boundary layer are investigated, respectively. The influence of the magnetic field on these factors is analyzed. The most significant first analytical solution will be the focus of further discussion.

\subsubsection{Velocity Distribution}
\begin{figure}
  \centerline{\includegraphics[width=0.6\textwidth]{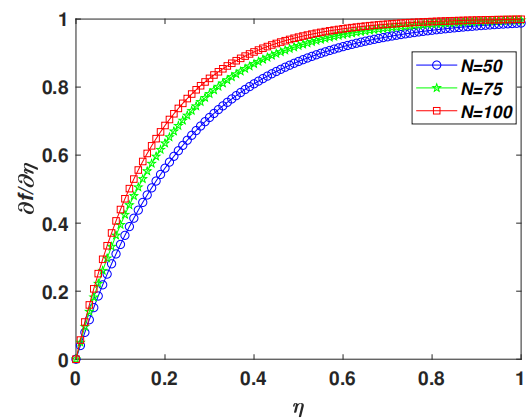}}
  \caption{Dimensionless velocity profiles under various parameters $\textit{N}$.}
\label{figu}
\end{figure}

Figure \ref{figu} illustrates the effect of the Stuart number on the velocity profile of the boundary layer. The overall velocity gradient of the boundary layer increases with an increase in the Stuart number. This observation signifies that an increase in the Lorentz force results in a reduction of the boundary layer thickness. Furthermore, the response of the boundary layer to the strength of the magnetic field demonstrates nonlinear characteristics.

\subsubsection{Friction and Thickness}
The damping effect on the wall, known as friction, caused by the viscosity of boundary layer flow, is an important physics problem that has received special attention. Based on the discussions above, it can be concluded that the normal gradient of flow velocity is influenced by the magnetic field, as indicated by the trend in boundary layer thickness. A quantitative and straightforward relationship needs to be derived to address this issue. Additionally, this relationship will help determine the extent to which magnetic field strength impacts friction. Newton's shear law gives the formula for the viscous stress at the wall as $\frac{\mathrm{d}F_d}{\mathrm{d}x}=\mu\frac{\partial u}{\partial y}\vert_{\eta=0}$. Here $F_d$ is the real damping force and the width perpendicular to the paper surface is taken as 1. The dimensionless friction density is defined as $f_d=\frac{\mathrm{d}F_d}{\mathrm{d}x}\frac{\delta}{\mu U}=\frac{\partial^2 f}{\partial\eta^2}\vert_{\eta=0}$. $f_d$ can be directly computed by the analytical solution of $g$ in equation (3.6)
\begin{equation}
    f_d=\frac{2^{-1/4}}{\sqrt{6}}(1+2N+2N^2)^{3/4}[\frac{\sinh{c_b}+\cosh{c_b}}{(\cosh{c_b})^3}]
\end{equation}

Note that as $N$ increases, the hyperbolic tangent of $c_b$ approaches 1. It is necessary to perform a first-order Taylor expansion on it to obtain a finite approximation $c_b\approx\tanh^{-1}(1-\frac{1}{4N})\approx\frac{1}{2}\ln{N}+\frac{1}{2}\ln{8}$. Substitute the approximate value of $c_b$ into the aforementioned equation and perform Taylor expansion once more to obtain:
\begin{equation}
    f_d\approx\frac{1}{\sqrt{3}}N^{1/2}
\end{equation}

\begin{figure}
  \centerline{\subfigure[]{\includegraphics[width=0.5\textwidth]{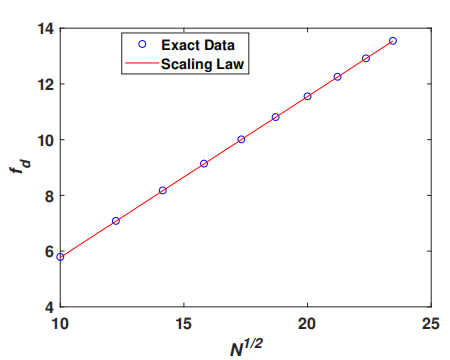}}\subfigure[]{\includegraphics[width=0.5\textwidth]{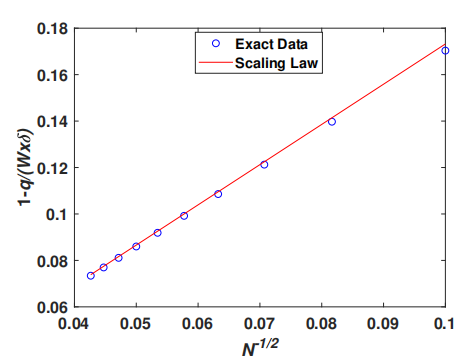}}}
  \caption{Scaling laws of friction (a) and flux (b) compared with analytical data.}
\label{figscaling}
\end{figure}

This represents a square root scaling law that illustrates the relationship between friction and magnetic field strength. Under the condition of the strong magnetic field, this scaling law closely matches the direct data obtained from the analytical solution, as shown in figure \ref{figscaling}(a). This also indirectly supports the earlier argument that a strong Lorentz force reduces the boundary layer thickness. When the magnetic field is strong, the average velocity gradient of the entire boundary layer is replaced by the velocity gradient at the wall, as $\frac{U}{\delta_B}\approx\frac{\partial u}{\partial y}\vert_{\eta=0}=f_d\frac{U}{\delta}$, substituting the above scaling law of $f_d$, the boundary layer thickness $\delta_B$ reduced by the strong Lorentz force can be estimated as
\begin{equation}
    \delta_B/\delta\approx\sqrt{3}N^{-1/2}
\end{equation}
The $\delta$ in the above equation can be estimated using the Reynolds number, as $\delta\approx L/\sqrt{Re}$. Stuart number is defined as $N=Ha^{2}/Re$, where $Ha$ is the Hartmann number. Therefore, it can be further transformed into $\delta_B/L\approx\sqrt{3}Ha^{-1}$. In the present model, the wall is located under the side layer, instead of the Hartmann layer. This seems to contradict the classical thickness of the side layer which is \textit{O}$(Ha^{-1/2})$ \citep{Schercliff1953, Hunt1965}. In fact, the Hunt's model only considered steady flow in a single direction and assumed that the main velocity remained constant along the flow direction. This resulted in the disappearance of both transient and convective terms in the N-S equation of Hunt's model. However, in the present model, the inertial force cannot be ignored, and the main flow gradient $W$ of the velocity affects the boundary layer, so $W$ is retained in the definition of $N$. It results in different scaling laws for the boundary layer thickness. The present model can be further expanded. If the direction of the magnetic field is changed to be perpendicular to the wall, the electrical potential no longer contributes to the Lorentz force. The coefficients of Lorentz force in momentum equations change from 2 to 1, causing the $N$ to change to $N/2$, ultimately resulting in the coefficient of boundary layer thickness scaling law becoming $\sqrt{6}$, but the power of $Ha$ remains at -1. This is consistent with the thickness magnitude of the Hartmann layer \citep{Hartmann1937}.

\subsubsection{Flux in Boundary Layer}
In mathematical terms, the above two sections discuss the dependence of the first and second derivatives of the dimensionless stream function $f$ on a parameter caused by the strong magnetic field. In this section, the zero derivative of $f$ (i.e. itself) will be the focal point of discussion. It is important to note that this quantity possesses physical significance, as it corresponds to the velocity flux in the boundary layer region. It is not difficult to derive the flux formula $q=\int_{0}^{1}U\frac{\partial f}{\partial\eta}\delta \mathrm{d}\eta=U\delta f_\delta$ using the definition of Stokes stream function. The $f_\delta$ is the value of $f$ at $\eta=1$ position. The application of the Taylor expansion method from the preceding section enables the derivation of an approximate formula for the flux under conditions of strong magnetic fields.
\begin{equation}
    q\approx Wx\delta(1-\sqrt{3}\frac{1}{\sqrt N})
\end{equation}

As shown in figure \ref{figscaling}(b), the flux scaling law is consistent with the direct output data of the analytical solution (taking $Wx=1$). Physically, the Lorentz force promotes mass transport in the boundary layer. When the magnetic field is relatively strong, the concise inverse square root scaling law can be used to estimate the boundary layer flux.
This section elucidates the physical impact of the Lorentz force on the boundary layer of a transient stagnation flow using exact mathematical formulae.

\section{Converging Flow $(k=-1)$}\label{Converging}
The boundary layer solutions for converging channel flow, as investigated by \cite{Pohlhausen1921}, \cite{Landau1976}, and \cite{Sun2024}, can also be extended to an unsteady MHD model. Consider placing a thin wire carrying a constant electrical current intensity, denoted as $I$, perpendicular to the flow direction at the cusp (the coordinate origin). According to the Biot-Savart law, the distribution of the magnetic field generated by the current is $B(x)=\frac{\mu_0I}{2\pi x}$. In this context, the Lorentz force term in the momentum equation does not explicitly include the x-coordinate. To derive an analytical solution, we will apply the short-time approximation proposed by \cite{Sun2024}. The mathematical properties of the simplified momentum equation align completely with equation (3.4).
\begin{equation}
  \frac{\partial^3f}{\partial\eta^3}-\frac{\partial^2f}{\partial\eta\partial\tau}-\beta(\frac{\partial f}{\partial\eta})^2+\beta=\frac{\sigma I^2\mu_0^2\beta}{2\pi^2\rho W}(\frac{\partial f}{\partial\eta}-1)
\end{equation}

Still assigning coefficient $\beta=1$. Using the same substitution method, just replace the coefficients of the nonlinear solution $g$ to form the analytical solution of the current equation. The analytical solution becomes
\begin{equation}
  f(\eta, \tau)=-m\eta-\sqrt{3(1+m)}\tanh\xi+2\sqrt{3(1+m)}\ln\cosh\xi+h(\tau)
\end{equation}
\begin{equation}
  \xi=\frac{1}{2}\sqrt{\frac{1+m}{3}}\eta+\frac{5}{6}(1+m)\tau+c_b
\end{equation}
\begin{equation}
  c_b=\tanh^{-1}[(\sqrt{4m^2+6m+2}-1-m)/(1+m)]
\end{equation}
Here, $Ha_I=\sqrt{\sigma/\mu}I\mu_0/(2\pi)$ is defined as equivalent Hartmann number which characterizes the effect of the induced magnetic field produced by the current $I$. Then $m=\frac{\sigma I^2\mu_0^2}{4\pi^2\rho W}$ is the equivalent Stuart number $N_I$.

Similarly, the square root scaling law also applies to the wall damping, thickness and flux of the converging flow boundary layer, except that the coefficient becomes four times as $f_d\approx(4/\sqrt{3})\sqrt{N_I}$, $\delta_{B}/\delta=(\sqrt{3}/4){N_I}^{-1/2}$, $q\approx Wx\delta(1-4\sqrt{3}{N_I}^{-1/2})$. Presently, the physical effects of the non-uniform magnetic field on the boundary layer of unsteady converging flow are represented by analytical solutions and scaling laws.

\section{Conclusions}
This work presents analytical solutions for two types of unsteady magnetohydrodynamic (MHD) boundary layer flows using a semi-explicit similarity transformation method. The results demonstrate the significant influence of Lorentz force on the multiple waves structure of the boundary layer flow. Additionally, the scaling laws are derived that reveal a nonlinear relationship between boundary layer thickness, friction, flux and magnetic field strength. A key finding is that stronger magnetic fields consistently lead to boundary layer thinning and increased friction. Overall, this work provides a deeper insight into the transient coupling mechanism by which MHD effects govern boundary layer dynamics.

\backsection[Acknowledgements]{The authors gratefully acknowledge support from the Natural Science Foundation of China (NSFC) under Grant Nos. 52176089 and U23B20110.}
\backsection[Declaration of interests]{The authors report no conflict of interest.}
\backsection[Author ORCIDs]{Jing-Yu Fu https://orcid.org/0009-0009-7446-0388. Ming-Jiu Ni https://orcid.org/0000-0003-3699-8370. Nian-Mei Zhang https://orcid.org/0000-0001-5400-9607.}

\bibliographystyle{jfm}


\end{document}